\documentclass[useAMS,usenatbib]{mn2e}
\usepackage{multirow}
\usepackage{epsfig}
\usepackage{subfigure}
\usepackage{url}

\makeatletter
\let\jnl@style=\rm
\def\ref@jnl#1{{\jnl@style#1}}

\def\aj{\ref@jnl{AJ}}                   
\def\araa{\ref@jnl{ARA\&A}}             
\def\apj{\ref@jnl{ApJ}}                 
\def\apjl{\ref@jnl{ApJ}}                
\def\apjs{\ref@jnl{ApJS}}               
\def\ao{\ref@jnl{Appl.~Opt.}}           
\def\apss{\ref@jnl{Ap\&SS}}             
\def\aap{\ref@jnl{A\&A}}                
\def\aapr{\ref@jnl{A\&A~Rev.}}          
\def\aaps{\ref@jnl{A\&AS}}              
\def\azh{\ref@jnl{AZh}}                 
\def\baas{\ref@jnl{BAAS}}               
\def\jrasc{\ref@jnl{JRASC}}             
\def\memras{\ref@jnl{MmRAS}}            
\def\mnras{\ref@jnl{MNRAS}}             
\def\pra{\ref@jnl{Phys.~Rev.~A}}        
\def\prb{\ref@jnl{Phys.~Rev.~B}}        
\def\prc{\ref@jnl{Phys.~Rev.~C}}        
\def\prd{\ref@jnl{Phys.~Rev.~D}}        
\def\pre{\ref@jnl{Phys.~Rev.~E}}        
\def\prl{\ref@jnl{Phys.~Rev.~Lett.}}    
\def\pasp{\ref@jnl{PASP}}               
\def\pasj{\ref@jnl{PASJ}}               
\def\qjras{\ref@jnl{QJRAS}}             
\def\skytel{\ref@jnl{S\&T}}             
\def\solphys{\ref@jnl{Sol.~Phys.}}      
\def\sovast{\ref@jnl{Soviet~Ast.}}      
\def\ssr{\ref@jnl{Space~Sci.~Rev.}}     
\def\zap{\ref@jnl{ZAp}}                 
\def\nat{\ref@jnl{Nature}}              
\def\iaucirc{\ref@jnl{IAU~Circ.}}       
\def\aplett{\ref@jnl{Astrophys.~Lett.}} 
\def\apspr{\ref@jnl{Astrophys.~Space~Phys.~Res.}}
\def\bain{\ref@jnl{Bull.~Astron.~Inst.~Netherlands}}
\def\fcp{\ref@jnl{Fund.~Cosmic~Phys.}}  
\def\gca{\ref@jnl{Geochim.~Cosmochim.~Acta}}   
\def\grl{\ref@jnl{Geophys.~Res.~Lett.}} 
\def\jcp{\ref@jnl{J.~Chem.~Phys.}}      
\def\jgr{\ref@jnl{J.~Geophys.~Res.}}    
\def\jqsrt{\ref@jnl{J.~Quant.~Spec.~Radiat.~Transf.}}
\def\memsai{\ref@jnl{Mem.~Soc.~Astron.~Italiana}}
\def\nphysa{\ref@jnl{Nucl.~Phys.~A}}   
\def\physrep{\ref@jnl{Phys.~Rep.}}   
\def\physscr{\ref@jnl{Phys.~Scr}}   
\def\planss{\ref@jnl{Planet.~Space~Sci.}}   
\def\procspie{\ref@jnl{Proc.~SPIE}}   

\makeatother

\newcommand {\apgt} {\ {\raise-.5ex\hbox{$\buildrel>\over\sim$}}\ }
\newcommand {\aplt} {\ {\raise-.5ex\hbox{$\buildrel<\over\sim$}}\ }

\title[X-ray spectroscopy of ESO 138-G001]{Deep  X-ray spectroscopy and imaging of the Seyfert 2 galaxy, ESO 138-G001}
\author[M. De Cicco, et al.]{M. De Cicco$^1$\thanks{E-mail: m.decicco89@gmail.com (MD)}, A. Marinucci$^1$, S. Bianchi$^1$, E. Piconcelli$^{2}$, G. Violino$^{3}$, 
\newauthor  C. Vignali$^{4,5}$, F. Nicastro$^{2,6,7}$\\
$^1$Dipartimento di Matematica e Fisica, Universit\`a degli Studi Roma Tre, via della Vasca Navale 84, 00146 Roma, Italy\\
$^2$INAF- Osservatorio Astronomico di Roma, Via di Frascati 33, 00040, Monte Porzio Catone, RM, Italy\\
$^3$Centre for Astrophysics Research, University of Hertfordshire, College Lane, Hatfield, AL10 9AB, UK\\
$^4$Dipartimento di Fisica e Astronomia, Universit\`a degli Studi di Bologna, Viale Berti Pichat 6/2, 40127 Bologna, Italy\\
$^5$INAF-Osservatorio Astronomico di Bologna, Via Ranzani 1, 40127 Bologna, Italy \\
$^6$ Harvard-Smithsonian Center for Astrophysics, 60 Garden Street, MS-04, Cambridge, MA 02138, USA\\
$^7$ Department of Physics, University of Crete, PO Box 2208, GR-710 03 Heraklion, Crete, Greece \\
}

\begin{document}

\maketitle

\label{firstpage}

\begin{abstract}
We present a spectral and imaging analysis of the XMM-{\it Newton} and {\it Chandra} observations of the Seyfert 2 galaxy ESO138-G001, with the aim of characterizing the circumnuclear material responsible for the soft (0.3-2.0 keV) and hard (5-10 keV) X-ray emission. We confirm that the source is absorbed by Compton-thick gas. However, if a self-consistent model of reprocessing from cold toroidal material is used (\textsc{MYTorus}), a possible scenario requires the absorber to be inhomogenous, its column density along the line of sight being larger than the average column density integrated over all lines-of-sight through the torus. The iron emission line may be produced by moderately ionised iron (FeXII-FeXIII), as suggested by the shifted centroid energy and the low K$\beta$/K$\alpha$ flux ratio. The soft X-ray emission is dominated by emission features, whose main excitation mechanism appears to be photoionisation, as confirmed by line diagnostics and the use of self-consistent models (\textsc{Cloudy}).
\end{abstract}


\section{Introduction}

The X-ray spectrum of highly obscured Seyfert galaxies is dominated by reflection components, originating from both cold and ionised circumnuclear matter \citep{matt00c}. In unobscured objects, these reflection components can be heavily diluted, often down to the point of invisibility. However in Compton-thick sources (where the absorbing column density along the line of sight exceeds N$_{\rm H}=\sigma_{\rm T}^{-1}\simeq1.5\times10^{24}$ cm$^{-2}$) the complete obscuration of the primary nuclear continuum permits us to have a clear view of these components. So far, most of our knowledge of circumnuclear reprocessing material is based on the brightest Compton-thick sources in the sky, like Circinus \citep[e.g.][]{matt99,Sambruna01b,mbm03}, NGC~1068 \citep[e.g.][]{kin02,matt04}, Mrk~3 \citep[e.g.][]{sako00b,bianchi05b,pp05} and NGC~424 \citep{mbm11}.

ESO138-G1 ( a Seyfert 2 galaxy  at $z$ = 0.0091)   exhibits a compact nucleus and a bright asymmetric, wedge-shaped circumnuclear zone of diffuse light resembling an ionization cone from the AGN \citep{munoz07}. 
Studying  X-ray data from ASCA, \citet{colbra00} found for this object a hard spectrum and a prominent Fe K$\alpha$ emission line, suggesting  that ESO 138-G1 is  a reflection dominated source, but statistically acceptable fits were also found with a partial-covering absorption column of N$_H$=2$\times$10$^{23}$ cm$^{-2}$.
 
Recently, an X-ray spectral analysis of this source has been performed by \citet{pico11}, studying two short XMM-{\it Newton} observations in 2007.
A soft excess component, characterized by the presence of several emission lines, was revealed below 2 keV.
Fitting above 3 keV with a power law,  a very flat slope was found ($\Gamma$ $\sim$ 0.35), with the presence of a prominent emission line around 6.4 keV and identified with the neutral iron K$\alpha$ fluorescence line. The large EW ($\sim$800 eV) of this feature indicates a heavy obscuration along the line of sight. This obscuration can be caused by a Compton-thin (transmission scenario) or Compton-thick (reflection scenario) screen of absorbing material which prevents the direct observation of most or all nuclear X-ray emission. These models provided a good fit to the data and appear  statistically equivalent; however the  equivalent width of the Fe K$\alpha$ line around 800 eV and  the low ratio of 2-10 keV to de-reddened [OIII] fluxes lead \citet{pico11} to suppose that ESO 138-G1 is a Compton-thick galaxy. Finally, it has been noted that the upper limits to the 15-150 keV flux provided by Swift/BAT and INTEGRAL/IBIS observations seem to exclude the presence of a direct view of the nuclear continuum even in the very hard X-ray band, implying a very high value for the column density, of the order of 10$^{25}$ cm$^{-2}$. 

In this paper, we present a new observational campaign of ESO 138-G001 performed with XMM-{\it Newton} and {\it Chandra} in 2013 and 2014, respectively.

\section{Observations and data reduction}
 
\begin{figure}
\epsfig{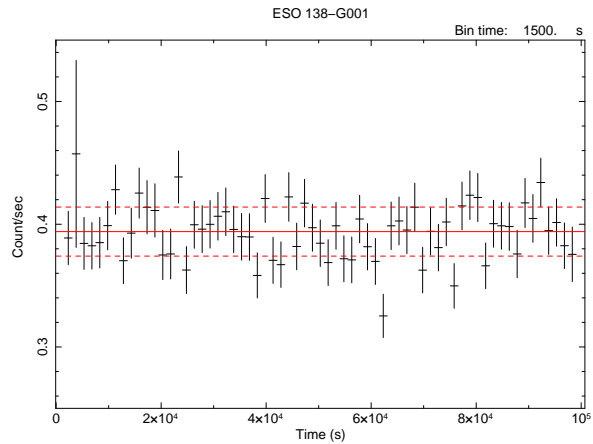}
\caption{Background-subtracted EPIC-pn light curve of the source in the 0.5--10 keV energy band. Horizontal solid and dashed lines indicate mean and standard deviation, respectively.}
\label{fig:lc}
\end{figure}

ESO 138-G001 was observed by \textit{XMM-Newton} \citep{xmm} twice in 2007 and once in 2013, on February 24, with the European Photon Imaging Cameras (EPIC): the pn \citep{struder01} and the two MOS  \citep[Metal-Oxide Semi-conductor;][]{turner01} detectors. These were operated in Full Frame Mode using thin filters. The 2007 observations, for a total clean exposure time of $\simeq27$ ks, are discussed in detail in \citet{pico11}. We focus here on the long observation performed in 2013, for total elapsed time of 135 ks. Data were reduced with SAS 14.0 with the latest calibration available at the time of writing. Screening for intervals of flaring particle background was done consistently with the choice of extraction radii, in an iterative process based on the procedure to maximize the signal-to-noise ratio described in detail by \citet{pico04} in their appendix A. This led to optimal extraction radii of 33 and 35 arcsec, respectively for pn and the two MOS, while background spectra were extracted from nearby circular regions with a radius of 50 arcsec. After having verified their consistency, the two MOS spectra were co-added. The final clean exposure times are 85 and 122 ks, for the pn and the co-added MOS spectra, respectively. Patterns 0--4 for pn and 0--12 for MOS were used. The spectra were then binned in order to over-sample the instrumental resolution by at least a factor of 3 and to have no less than 30 counts in each background-subtracted spectral channel, allowing us to adopt the $\chi^2$ statistics. Although we found no significant difference between the 2003 and the 2013 spectra, we decided not to co-add them, since the new observation has a much longer exposure time.  We also used time-averaged spectra, due to the lack of both spectral and flux variations in the 2013 observation (Fig. \ref{fig:lc}). Moreover, we do not use MOS data in this paper, unless explicitly stated. Finally, RGS spectra were extracted with standard procedures. Background spectra were generated using blank field event lists, accumulated from different positions on the sky vault along the mission. Exposure times of the RGS spectra are 128 ks (see Table~\ref{tab:Timesandcounts}).

The source was then observed by {\it Chandra} on 2014, June 20 for an exposure time of 49 ks, with the ACIS-S detector \citep[Advanced CCD Imaging Spectrometer:][]{acis}. Data were reduced with the Chandra Interactive Analysis of Observations \citep[CIAO:][]{ciao} 4.7 and the Chandra Calibration Data Base 4.6.5 data base, adopting standard procedures. The imaging analysis was performed applying the subpixel event re-positioning and smoothing procedures discussed in the literature \citep{tmm01, lkp04}.  We therefore used a pixel size of 0.246 arcsec, instead of the native 0.495 arcsec. At the distance of the source the adopted pixel size corresponds to $\simeq50$ pc. After cleaning for background flaring events, we get a clean exposure of 49 ks. Source and background spectra were extracted from circular regions of 5 and 15 arcsec and grouped them to have at least 20 total counts per new bin. The source is affected by pile-up and, following \citet{davis01}, we estimate it to be $\sim 10\%$. 

In the following, errors and upper limits correspond to the 90 per cent confidence level for one interesting parameter, where not otherwise stated. The adopted cosmological parameters are $H_0=70$ km s$^{-1}$ Mpc$^{-1}$ , $\Omega_\Lambda=0.73$ and $\Omega_m=0.27$ \citep[i.e. the default ones in \textsc{xspec 12.8.1}:][]{xspec}. All models presented in this paper include Galactic absorption along the line of sight \citep[N$_H$=1.3$\times$10$^{21}$cm$^{-2}$: ][]{kalberla05}, i.e. the same value used by \citet{pico11}.

\section{X-ray spectral analysis}
\subsection{\label{hard} The hard X-ray (5-10 keV) band}

As a first step, we restrict our analysis to the 5-10 keV energy range to investigate the reflected primary continuum and the iron K$\alpha$ complex, testing two alternative scenarios:
\begin{itemize}
\item  a reflection-dominated scenario where the X-ray primary continuum from the AGN is totally suppressed in the EPIC band for the presence of a Compton-thick absorber ($\mathrm{N_H}>1.5\times10^{24}$cm$^{-2}$).
\item  a transmission-dominated scenario,  where the AGN emission is absorbed by a Compton-thin ($\mathrm{N_H}<1.5\times$10$^{24}$ cm$^{-2}$) obscuring screen.
\end{itemize}  

We start fitting the pn data  with the reflection scenario. This model consists of a neutral absorption component modeled by \textsc{TBABS}, using solar
abundances \citet{wilm01} and cross-sections from \citet{verner96}, and a cold reflection component modeled by \textsc{pexrav} \citep{mz95} in \textsc{xspec}. The reflection factor $R$ indicates the amount of radiation reflected by Compton scattering from a semi-infinite slab with respect to the incident continuum. We assume solar abundances, $R=-1$ (corresponding to a pure reflection spectrum) and we fix the high-energy cutoff to E$_c$=1000 keV and the inclination angle to $\theta=65^{\circ}$ \citep{pico11} . Additional free parameters are the photon index $\Gamma$
and four Gaussian emission lines (see section~\ref{emissionlines} for further details) corresponding to the strong fluorescence Fe K$\alpha$ and Ni K$\alpha$ lines, the Fe K$\beta$ and an ionised Fe K$\alpha$ line (Fe XXV K$\alpha$), as already reported in \citet{pico11}. We fix all the line widths $\sigma$ to zero, exception made for the Fe K$\alpha$. Fitting the pn data with the reflection model,  we obtain $\chi^{2}$/dof=78/78=1 with a photon index  $\Gamma$=1.8$\pm 0.1$. In Fig. 1  we show the pn data set fitted with the reflection model and residuals (in black).

\begin{figure}
\begin{center}
\epsfig{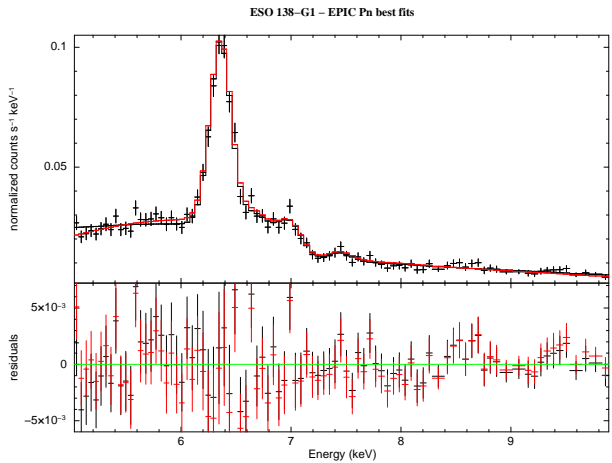}
\caption{Reflection (in black) and transmission (in red) models fitted to pn data are shown, with residuals in the 5-10 keV energy range.}
\end{center}
\label{fig:pnreflection}
\end{figure}

\begin{table*} 
\begin{center}
\begin{tabular}{cccccc}
\hline 
Line & Energy  & EW & Intensity & $\sigma$ & E$_T$ \\ 
 & (keV) &(eV) & $\times$10$^{-6}$(ph cm$^{-2}$ s$^{-1}$) & (eV)& (keV)\\
\hline 
& & & & & \\
Fe K$\alpha$ & 6.417$^{+0.007}_{-0.006}$ & 745$\pm 45$ & 26.7$\pm 1.7$  & 22$^{+25}_{-20}$ & 6.400 \\
& & & & \\ 
\multirow{2}{*} {Fe XXV K$\alpha$} &\multirow{2}{*} {6.73$\pm 0.07$} &\multirow{2}{*} {54$\pm 30$} &\multirow{2}{*} {2.0$\pm 1.0$} &\multirow{2}{*} {-}& 6.675(i) \\ & & & & & 6.700(r)\\
& & & & \\
Fe K$\beta$ & 7.07$\pm 0.04$ & 100$\pm 30$ & 3.1$\pm 0.8$  & - & 7.058 \\ 
& & & & \\
Ni K$\alpha$ & 7.51$\pm 0.07$ &  80$\pm 40$ & 1.6$\pm 0.8$ & - & 7.472 \\ 
\hline
\end{tabular}
\end{center}

\caption{\label{tab:hardlines} Best fit parameters for the 5-10 keV pn analysis. Observed centroid energies, Equivalent Widths, intensities, widths and theoretical transition energies are reported.  }
\end{table*}

\begin{figure}
\begin{center}
\epsfig{file=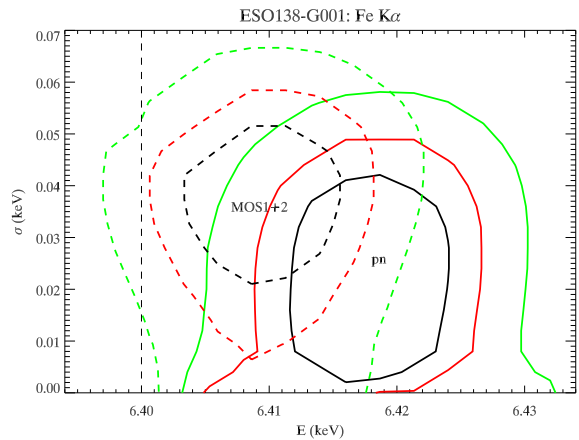, width=\columnwidth}
\caption{Contour plots between the width and the energy centroid of the Fe K$\alpha$ line are shown for both pn and MOS data (solid and dashed lines, respectively). Each curve represents a confidence level: black curve 68$\%$, red curve 90$\%$, green curve 99$\%$. Vertical dashed line indicates the theoretical transition energy.}
\end{center}
\label{fig:sigmaFekalpha}
\end{figure}

\begin{table}
\begin{center}
\begin{tabular}{c c c c}
\hline
& date & exposure time & count rates\\
& & (ks) & cts/s\\
\hline
RGS 1 & 2013-02-24 &128 &(6.1$\pm$0.8)$\times10^{-3}$ \\
RGS 2 & 2013-02-24 &128 & (8.1$\pm$0.7)$\times10^{-3}$ \\
pn & 2013-02-24 & 85& 0.39$\pm$0.02 \\
ACIS & 2014-06-20 & 49& (9.8$\pm$0.1)$\times10^{-2}$ \\
\end{tabular}
\end{center}
\caption{\label{tab:Timesandcounts} Dates, exposure times and count rates. For the two RGS the energy range is 0.3-2.5 keV, while for pn and ACIS the energy range is 0.5-10 keV. }
\end{table}

 As a next step we fit the pn data with the transmission scenario. The transmission model is composed by an absorbed  power-law (in which the Galactic absorption is modeled by \textsc{TBABS}) and four Gaussians to reproduce the four emission lines discussed above. Since no reflection from cold material is considered we do not use the \textsc{pexrav}
component in our model. We get a $\chi^{2}$/dof=79/77=1.02 with $\Gamma$=2.6$\pm0.3$ and N$_{\rm H}=5.8^{+0.6}_{-1.2}\times10^{23}$ cm$^{-2}$. We plot the data set fitted with the transmission model in Fig. 1 and residuals (in red).

\begin{table}
\begin{center}
\begin{tabular}{c c c}
\hline
Band & Flux & Luminosity \\
(keV)& $\times$10$^{-12}$ erg cm$^{-2}$ s$^{-1}$&$\times$10$^{41}$ erg s$^{-1}$\\
\hline
5-10  & 1.89$^{+0.03}_{-0.08}$  & 3.46$^{+0.06}_{-0.13}$ \\
2-10 & 2.29$^{+0.04}_{-0.09}$  & 4.20$^{+0.06}_{-0.17}$\\
0.5-10  & 2.33$^{+0.03}_{-0.10}$ & 4.26$^{+0.07}_{-0.17}$ \\
\hline

\end{tabular}
\end{center}
\caption{\label{tab:fluxluminosities}Observed fluxes and luminosities. }
\end{table}

From a statistical point of view, the reflection model and the transmission model appear equivalent (reduced $\chi^2$ values of 1 and 1.02, respectively, for one additional degree of freedom)but for the reflection model the best fit value for the   the power law index is typical of Seyfert galaxies \citep{bianchi09} and it is in agreement with the value reported in \citet{pico11}. In the transmission scenario, the power law index is extremely steep ($\Gamma$=2.6$\pm0.3$) and cannot be reconciled with the observed distribution of this parameter in radio-quiet AGN. Moreover, we can use the ratio $T$ of the observed 2-10 keV  to the extinction-corrected [OIII] fluxes as a proxy for the amount of obscuration of the X-ray primary continuum. The [OIII] emission line is an isotropic indicator of the AGN power, being produced in the Narrow Line Region. In particular, X-ray sources with $T<1$ are associated with a Compton-thick absorber and a 2-10 keV reflection dominated spectrum \citep{bass99, akygeo09, gm09}. The extinction-corrected [OIII] flux of ESO 138-G1 is 2.7$\times$10$^{-12}$erg cm$^{-2}$ s$^{-1}$ (\citet{pico11} and references therein). In our reflection scenario $T$ assumes a value (with the 2-10 keV flux reported in  Table 2) of $\sim$ 0.85, which is consistent with the existence of a Compton-thick screen which suppresses the intrinsic 2-10 keV flux. For the transmission scenario, we found for the de-absorbed 2-10 keV flux a value of 1.4$\times$10$^{-11}$erg cm$^{-2}$ s$^{-1}$, which gives  T  $\sim$5. This value is inconsistent with those found by many studies for unabsorbed Seyfert 1 and absorption corrected Compton-thin Seyfert 2 galaxies \citep{mai98, panessa06, lamastra09}, which are typically $\geq$ 10. Therefore, the comparison with the [OIII] flux gives further support to the reflection dominated scenario.

\subsubsection{\label{emissionlines} The emission features}

All the line fluxes (see Table~\ref{tab:hardlines}) are consistent with the values found by \citet{pico11} for the old 2007 XMM observation. We find an intense Fe K$\alpha$ emission line at 6.417$^{+0.007}_{-0.006}$ keV: its large Equivalent Width (EW=745$\pm45$ eV) suggests the presence of Compton-thick material \citep{mpp91,gf91}. Fe K$\beta$ and Ni K$\alpha$ are unresolved, but their width is consistent with being produced by the same gas responsible for the Fe K$\alpha$ emission. We find an emission line at 6.73$\pm 0.07$ keV, consistent with being dominated by contributions from the resonance and the inter-combination components of the Fe XXV K$\alpha$ triplet.
 
 The energy centroid of the Fe K$\alpha$ line \citep[consistent with the value found by][]{pico11} is not consistent  with the theoretical value for neutral iron in the pn spectrum, only marginally in the MOS (at a confidence level of 99$\%$, see Fig. 3). This inconsistency can be due to calibration issues, but may also have a physical origin, suggesting that iron may be ionised: the observed centroid energy is indeed consistent with FeXIII-FeXVI \citep{house69}.
We also calculated the ratio between the fluxes of Fe K$\beta$ and Fe K$\alpha$, which depends on iron ionization:

\begin{equation}
\frac{F_{FeK\beta}}{F_{FeK\alpha}}=0.116\pm{0.033},
\label{eq:Fekbeta/Fekalpha}
\end{equation}

This value again is inconsistent with the one expected for neutral iron \citep[0.155-0.160:][]{mol03, house69}, but it is consistent with the values of the ratio associated to the group of lines FeX-FeXII \citep{km93}. Although this piece of information is not formally in complete agreement from what we derived from the centroid energy, both diagnostics seems to agree that iron is ionised at about FeXII-XIII.

Adopting the best fit centroid energy and width, we find that FWHM=2400$^{+2500}_{-2100}$ km$s^{-1}$. From the FWHM, assuming a Keplerian motion around the central object, we can calculate the distance of the gas which produces the line:
\begin{equation}
FWHM=2v_k \sin i\simeq 1300 \left(\frac{M_8}{r}\right)^{1/2} \sin i \quad \rm  km\ s^{-1}
\label{eq:FWHM-mass}
\end{equation}
where the radius is expressed in parsec, the mass in 10$^8$M$_\odot$ and \textit{i} is the angle between the torus axis and the line of sight \citep[see e.g.][]{bianchi05b}.
The BH mass of ESO 138-G1 is estimated to be $\sim$4.6$\times$10$^{6}$M$_\odot$ \citep{pico11}. Inserting this value for different choices of i of 30 or 60 degrees we obtain respectively for the distance of the gas the values 3.4$\times$10$^{-3}$pc and 1$\times$10$^{-2}$pc. These values, together with the measured FWHM, are more suggestive of an origin of the line in the BLR rather than the torus. However, since the X-ray emission of this source is strongly absorbed, an ad-hoc geometry would be needed to explain the observation of an iron emission line from the BLR.

On theoretical grounds, the Compton-scattering of the red part of the Fe K$\alpha$ line should produce a Compton Shoulder, which can be modeled with a Gaussian line with $\sigma$ fixed at 40 eV and energy fixed at 6.3 keV \citep{matt02}.
The inclusion of such a component  produces only a little improvement of $\chi^2$ ($\chi^{2}$/dof=76/77=0.99) and we only find an upper limit for its intensity F$_{\rm CS}<4.3\times$10$^{-6}$ph cm$^{-2}$s$^{-1}$. The intensity and energy centroid of the iron K$\alpha$ are consistent with the values reported above and we find an upper limit of 36 eV for the width of the line. The ratio F$_{\rm CS}$/F$_{\rm Fe\ K\alpha}<$0.19 is marginally consistent with the 0.2 value expected for a Compton-thick material \citep{matt02, ym11}.

\subsection{The soft X-ray (0.3-2.5 keV) band}

The soft X-ray spectrum of ESO 138-G1 appears dominated by line emission, as commonly found in this class of sources \citep[e.g.][]{gb07}. We analyse the RGS 1 and RGS 2 spectra jointly, performing phenomenological fits on $\simeq100$-bin spectral segments, using Gaussian profiles at the redshift of the source (z=0.0091), and a power law, both absorbed by the Galactic column density along the line of sight.  Since the model used to fit the continuum is not very sensitive to the photon index $\Gamma$, due to the very limited band width of each segment, it has been fixed to $1$. Emission lines from H-like and He-like O, Ne, Mg and Si, as well as from the Fe L-shell, are detected (see Table~\ref{rgslines}). Even though the likely identification of RRC features, together with the dominance of OVII K$\alpha$ line, suggest an origin of these lines in a gas in photoionisation equilibrium \citep{gb07}, it is impossible to extract more pieces of information from this phenomenological fit.

\begin{figure}
\begin{center}
\epsfig{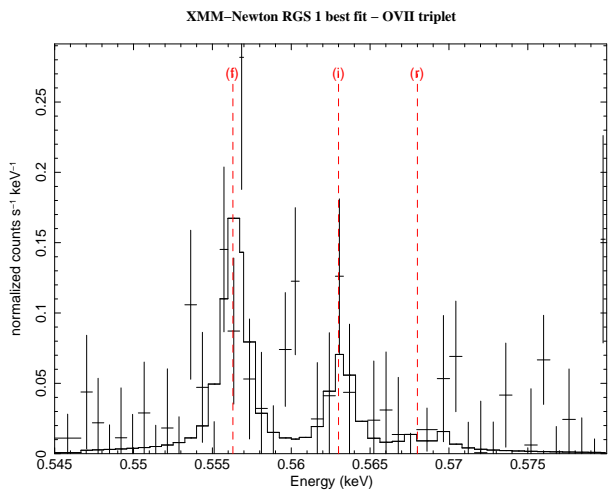}
\caption{\label{fig:ovii} Best fit model to the O VII K$\alpha$ triplet, using the RGS 1 spectrum.}
\end{center}

\end{figure}

\begin{table*}
\begin{center}
\begin{tabular}{c c c c c c}
\hline
Energy& Intensity& Energy & Intensity & Identification&E$_T$\\
(pn)& (pn)& (RGS)& (RGS)& & \\
\hline
& & & & &\\
\multirow{3}{*}{0.57$\pm 0.02$}&\multirow{3}{*}{28$^{+21}_{-11}$} & 0.5614$\pm 0.0004$  &  39$^{+19}_{-10}$ &\multirow{3}{*}{OVII K$\alpha$} &0.5610(f)\\
&  & 0.5685$\pm 0.0035$& 20$\pm13$& &0.5685(i)\\
& & 0.5739 *& $<38$ & &0.5739(r)\\
& & & & &\\
0.68$\pm 0.04$ &8.5$^{+7.7}_{-4.3}$ & 0.6530$\pm 0.0007$&18.4$\pm0.8$ &OVIII K$\alpha$ &0.6536\\
& & & & &\\
\multirow{2}{*}{0.78$^{+0.07}_{-0.15}$}&\multirow{2}{*}{15$^{+3}_{-7}$}&0.7263$\pm 0.018$&4.5$\pm 2.5$&FeXVII M2, FeXVII 3G&0.7252, 0.7272\\
&   & 0.7390$\pm 0.0005$&14.0$\pm 0.5$&OVII RRC & 0.7393\\
& & & & &\\
0.8257 * &$<$17 & 0.8263$^{+0.0015}_{-0.0008}$ & 7.5$\pm3.0$  & FeXVII 3C &0.8257\\
& & & & &\\
0.87$^{+0.06}_{-0.17}$&22$^{+16}_{-14}$ & 0.8738$^{+0.0002}_{-0.0012}$ &  12.0$\pm3.0$   & OVIII RRC,Fe XVIII L  &0.8714, 0.8728\\
& & & & &\\
\multirow{3}{*}{0.94$\pm 0.04$}&\multirow{3}{*}{24$^{+10}_{-12}$} & 0.9054$\pm 0.0005$  & 12.3$\pm0.4$ &   \multirow{3}{*}{NeIX K$\alpha$}&0.9050(f)\\
& & 0.9149 *&$<0.5$ & & 0.9149(i)\\
& & 0.9220$^{+0.006}_{-0.024}$&10.0$^{+4.7}_{-3.8}$ & &0.9220(r)\\
& & & & &\\
1.06$\pm0.02$&9.2$^{+2.0}_{-1.6}$ & 1.0747$^{+0.0007}_{-0.0017}$ & 5.8$\pm 2.7$ & NeIX K$\beta$ &1.0737 \\
& & & & &\\
1.19$\pm 0.01$ &8.7$\pm 1.5$ & 1.202$\pm 0.003$& 8.7$^{+1.7}_{-2.2}$& Fe XXIII L&1.170\\
& & & & &\\
\multirow{3}{*}{1.31$\pm 0.02$}&\multirow{3}{*}{8.3$\pm 1.6$} & 1.331$^{+0.012}_{-0.013}$&3.8$\pm2.3$ &\multirow{3}{*}{MgXI} &1.3311(f)\\
& & 1.3433 *&$<3$ & & 1.3433(i)\\
& & 1.351$^{+0.010}_{-0.026}$ & 4.3$^{+2.2}_{-1.7}$ &  &1.3522(r) \\
& & & & &\\
1.41$\pm 0.02$&5.7$^{+7.3}_{-4.4}$ & 1.4857$^{+0.047}_{-0.060}$ & 2.8$\pm 1.5$ & MgXII K$\alpha$ & 1.4723 \\
& & & & &\\
\multirow{3}{*}{1.82$\pm 0.01$}&\multirow{3}{*}{5.9$\pm 0.6$} & 1.8394 * &$<4.1$ &\multirow{3}{*}{SiXIII}  &1.8394(f) \\ 
 &  & 1.8541 * &$<4.8$ & & 1.8541(i)\\
& & 1.8649 * &$<5.8$ & & 1.8649(r)\\
& & & & &\\ 
1.98$\pm 0.04$&1.0$\pm 0.6$ & 2.0054 * &  $<13$     &Si XIV K$\alpha$  & 2.0054\\
& & & & &\\
2.41$\pm 0.02$&3.4$\pm 1.7$ & 2.416 *& $<30$& S XIV K$\alpha$&2.416\\
& & & & &\\
\hline
\end{tabular}
\end{center}
\caption{\label{rgslines}Emission lines detected in the soft X-ray (0.5-2.5 keV) analysis. We report their centroid energy (keV), intensity, identification and theoretical energy (keV). Intensities are in $10^{-6}$ ph cm$^{-2}$ s$^{-1}$ units, energies are in keV units. Asterisks indicate fixed energies.}
\end{table*}

\begin{figure}
\begin{center}
\epsfig{file=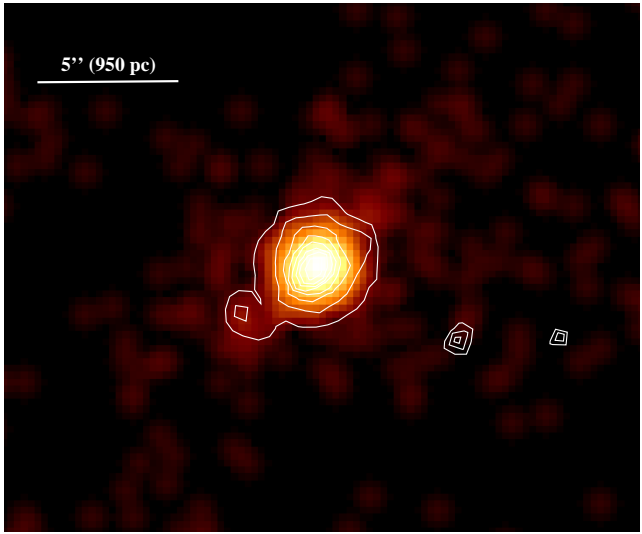, width=\columnwidth}
\caption{\label{chandra1}ACIS-S image of ESO138-G1, filtered between 0.3 and 2 keV. HST-WFPC2 contours (using [OIII] filter FR533N) are overimposed. North is up and East to the left.}
\end{center}
\end{figure}

In Fig.~\ref{chandra1} we show the 0.3-2 keV \textit{Chandra} ACIS image of ESO 138-G001. The source appears unresolved, but presents hints of elongation along the NW-SE direction: all the soft X-ray emission, likely arising from the Narrow Line Region, is enclosed in $\sim$1 squared kiloparsec. \citet{schmberg95} found that the [OIII] emission exhibits a jetlike structure that propagates westward up to $\sim$2 kpc from the nucleus at the 2$\sigma$ level, with a peak of emission at 7 arsec ($\sim$1.3 kpc). This [OIII] extension is also present in a HST/WFPC2 image presented by \citet{ferruit00}, where it appears as a faint extension given the short exposure. However, only 4\% of the total [OIII] flux of the source ($9.72\times 10^{-13}$ erg cm$^{-2}$ s$^{-1}$) is emitted outside the central 5 arcsec \citep[Table 2;][]{schmberg95}.  The soft X-ray flux of the NLR is generally a factor 2-5 lower than the [OIII] flux \citep{bianchi06}, so it is not surprising that this further weak extension is not detected in our {\it Chandra} observation.

\subsection{The broad-band (0.5-10 keV) model}

We then analyse the 0.5-10 keV spectrum of ESO 138-G1, by fitting the pn data with a new model consisting of the reflection model presented above, plus an unabsorbed power law component and a sequence of narrow ($\sigma$=0) Gaussian lines to account for the emission features observed in the soft part of the energy  range. We obtain a reasonable fit with  a reduced  $\chi^2$/dof=221/194=1.14. The unabsorbed power law component is described by a photon index $\Gamma$=2.80$\pm 0.03$, which is  consistent with the value found by \citet{pico11}. The reflection component is now described by a photon index $\Gamma$=1.59$^{+0.07}_{-0.04}$, which is marginally consistent with the photon index found in the hard band analysis $\Gamma$=1.8$\pm 0.1$.  In our analysis  we identify thirteen emission lines in the soft X-ray band, some of which already reported by \citet{pico11}. In Table~\ref{rgslines}, the best fit parameters of these lines are listed, together with a comparison with the RGS analysis presented in the previous section. No significant differences, both in energies and intensities, can be seen between the results of the two instruments.
We plot in Fig.~\ref{fig:broadbandfit} (left-hand panel) our data fitted with the broad-band model, which is shown in detail, with all its components, in Fig.~\ref{fig:broadbandmodel}. Observed fluxes and luminosities for this model in different energy bands are listed in Table~\ref{tab:fluxluminosities}.

We use the pn best-fit model to model the ACIS spectrum of the source, adding the \textsc{pileup} component in \textsc{xspec} \citep{davis01}, since the observation is affected by pile-up. Leaving as free parameters only the overall normalization, and the centroid energies and intensities of the Iron K$\alpha$ and Iron XXV K$\alpha$ emission lines, we get a best-fit with $\chi^2$/dof=199/176=1.13. All free parameters are in agreement with the ones presented above for the pn. We show in Fig.~\ref{fig:broadbandfit} (right-hand panel) the best-fit and residuals.
 
\begin{figure*}
\begin{center}
\epsfig{file=broadbandfit.ps, angle=-90, width=\columnwidth}
\epsfig{file=eso138_fitACIS.ps, angle=-90, width=\columnwidth}
\caption{\label{fig:broadbandfit}Broad band model fitted to pn (left) and ACIS (right) data with residuals in the 0.5-10 keV energy range.}
\end{center}

\end{figure*}

\begin{figure}
\begin{center}
\epsfig{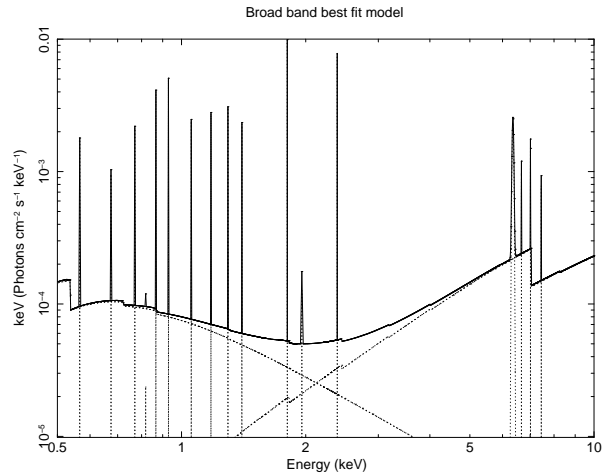}
\caption{\label{fig:broadbandmodel}Plot of the broad band model in the energy range 0.5-10 keV.}
\end{center}

\end{figure}

\section{A Self-consistent model}

As a final step, we performed a self-consistent modeling of the broad-band pn data, following the phenomenological analysis. In the previous sections we modelled the reflection emission from the cold material in ESO 138-G1 with \textsc{pexrav} which assumes a plane slab of cold reflecting material.  For the high energy part of the spectrum, we now adopt (as well as \textsc{TBABS} to model the Galactic column) \textsc{MyTORUS}\citep{my09}, which reproduces the primary continuum together with the reprocessed emission from a toroidal distribution of material. This scenario corresponds to the doughnut type of geometry used for the obscuring torus in the AGN unification schemes. We define the equatorial column density $N_H$ as the equivalent hydrogen column density through the diameter of the tube of the torus.

We used all three \textsc{MyTORUS} components: the primary power-law absorbed by the torus along the line of sight, the continuum reflection component and the emission features associated to it (Fe K$\alpha$, Fe K$\beta$ and the Fe K$\alpha$ Compton Shoulder). The normalization between these three components is left free to vary, in order to account for deviation from the standard geometry, covering factor and elemental abundances. On the other hand, the equatorial column density and the inclination angle of the torus are the same for all the components. Since the Fe K$\alpha$ line in ESO~138-G1 was found to have an energy significantly larger than 6.4 keV (see Sect.~\ref{hard}), we left the redshift free for the emission features component.

For the soft X-ray emission, we used an updated and extended version of the same photoionisation model first presented in \citet{bianchi10}, produced with \textsc{cloudy} 13.03 \citep[last described in][]{ferl00}. The main ingredients are: plane parallel geometry, with the flux of photons striking the illuminated face of the cloud given in terms of ionisation parameter U \citep{of06}; incident continuum modeled as in \citet{korista97}\footnote{Since ESO~138-G1 is a strongly absorbed source, we have no direct information on its intrinsic SED. However, the standard AGN SED that we used for this model is well suited for our qualitative analysis of a low resolution soft X-ray spectrum.}; constant electron density n$_e=10^5$ cm$^{-3}$ (we are nonetheless in a regime insensitive to density: \citet{pd00}); elemental abundances as in Table 9 of \textsc{cloudy} documentation. Only the reflected spectrum, arising from the illuminated face of the cloud, has been taken into account in our model.

\begin{figure}
\begin{center}
\epsfig{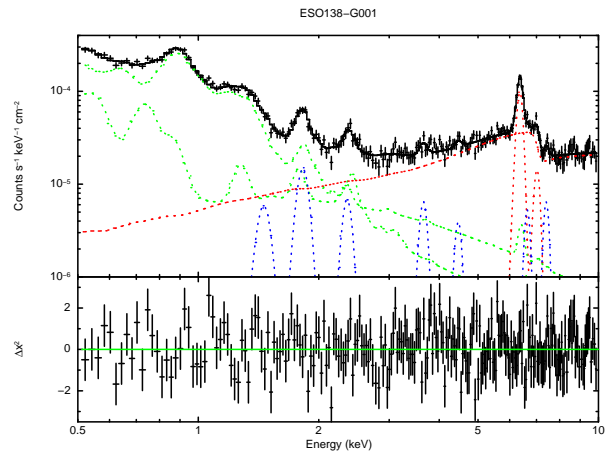}
\caption{\label{fig:mytorus} The self-consistent model described in Sect. 4 applied to the XMM-\textit{Newton} pn data in the energy range 0.5-10 keV. The spectrum (divided by the effective area) is shown in the upper panel, together with the total best fit (black line), and all the components: \textsc{MyTORUS} scattered components and lines (red), the two \textsc{cloudy} photoionisation components (green), and the seven Gaussian emission lines (blue). The residuals in terms of $\Delta\chi^2$ deviations are shown in the lower panel.}
\end{center}

\end{figure}

The resulting fit is unacceptable ($\chi^2=437/222$ d.o.f.), with strong residuals, mostly in the soft X-ray part of the spectrum. While adding emission from a collisionally ionised gas (\textsc{apec}) does not improve the fit significantly ($\Delta\chi^2<11$), another photoionisation component is strongly required by the data ($\chi^2=302/219$ d.o.f.). Some residuals are still present in the data: they can be readily identified with emission of fluorescent lines from neutral Nickel (7.472 keV), Calcium (3.690 keV), and Titanium (4.510 keV), which may be self-consistently produced in the torus, but are not included in \textsc{MyTORUS}, the FeXXV K$\alpha$ line (marginally constrained to be the resonant component of the triplet at 6.700 keV), not associated in our model to any continuum emission (the ionization parameters of both photoionisation components are too low to produce a significant contribution to this line: see below), as well as contribution from the forbidden transition of the SIXIII K$\alpha$ triplet (1.8394 keV), SiXIV K$\beta$ (2.3763 keV), and MgXII K$\alpha$ (1.4723 keV), which are underestimated by the two photoionisation components. The inclusion of these seven Gaussian lines  further improves the fit, which is now good ($\chi^2=239/212$ d.o.f.), and without systematic residuals (see Fig. ~\ref{fig:mytorus}).  All the energies of the Gaussians are fixed and their widths are fixed to be narrow ($\sigma$ = 0).
 The final model, in {\textsc xspec}, reads as:\\

{\textsc TBABS$\times$(CLOUDY$_1$+CLOUDY$_2$+7$\times$ZGAUSS\\
+POW$\times$MYT$_{\rm T}$+MYT$_{\rm S}$+MYT$_{\rm L}$).\\
}

The best fit parameters for this fit are summarized in Table~\ref{torusparameters}. The two photoionisation components have $\log \mathrm{U_1}=-0.80\pm0.05$, $\log \mathrm{U_2}=1.62^{+0.04}_{-0.03}$, and $\log \mathrm{N_{H1}}=22.4\pm0.1$, $\log \mathrm{N_{H2}}=21.13\pm0.12$ (in cm$^{-2}$). The photon index of the intrinsic continuum is $\Gamma=2.03\pm0.14$ and the equatorial column density for all  
\textsc{MyTORUS} components is $7.7\pm0.16\times10^{23}$ cm$^{-2}$. The inclination angle is constrained to be $<68\,^{\circ}$ (99\% confidence level), while the default geometry of \textsc{MyTORUS} requires it to be $>60\,^{\circ}$ in order to have absorption along the line of sight. The normalization between the emission lines component and the scattered continuum from the torus is $0.61^{+0.07}_{-0.05}$, reflecting a somewhat different geometry, covering factor and/or elemental abundance with respect to the default one of the model. 

\begin{table}
\begin{center}
\begin{tabular}{c c  }
\hline 
 Parameter &  Best fit value   \\ 
 \hline
&  \\ 
$\log \mathrm{U_1}$ & $-0.80\pm0.05$ \\
&\\
$\log \mathrm{NH_1}$ & $22.4\pm0.1$ \\
& \\
$\log \mathrm{U_2}$ & $1.62^{+0.04}_{-0.03}$ \\
& \\
$\log \mathrm{NH_2}$ & $21.13\pm0.12$\\
& \\
N$_H$ &  $7.7\pm0.16$    \\ 
&   \\
$\Gamma$ & $2.03\pm0.14$   \\
&  \\
MyTorusL/MyTorusS & $0.61^{+0.07}_{-0.05}$ \\
&  \\
MgXII K$\alpha$ (1.4723 keV) & $0.9\pm0.7$\\
& \\
SIXIII K$\alpha$ f (1.8394 keV) & $2.3\pm0.6$\\
& \\
SiXIV K$\beta$ (2.3763 keV) & $1.3\pm0.8$\\
& \\
CaI K$\alpha$ (3.690 keV) & $1.2\pm0.6$\\
&\\
TiI K$\alpha$  (4.510 keV) & $0.8\pm0.6$\\
&\\
FeXXV K$\alpha$ r (6.700 keV) & $1.3\pm1.0$\\
& \\
NiI K$\alpha$ (7.472 keV) & $1.7\pm0.8$\\
& \\
\hline
\label{torusparameters}
\end{tabular}
\end{center}
\caption{ Best fit values for the \textsc{MyTORUS} modeling. Intensities are in $10^{-6}$ ph cm$^{-2}$ s$^{-1}$ units, energies are in keV units and column density is in 10$^{23}$ cm$^{-2}$ units. }
\end{table}

Finally, the data do not require any contribution from a transmitted primary continuum, being it constrained to be $<10\%$ than the scattered component (99\% confidence level). The geometry and/or covering factor of the reflector may be significantly different from the one assumed by \textsc{MyTORUS}, greatly enhancing the reflection flux, or the observed reflected flux is an echo of a much more luminous state of the source in the past. Interestingly, \citet{rec15}, using WISE mid-IR colours (4.6--22 $\mu$m), suggested a torus inclination angle of $\sim 35^{\circ}$ for this source: in agreement with \citet{gab12}, in which the lack of $9.7\mu$m silicates absorption feaures is reported. Such a low inclination angle would support a scenario whereby we have an unobscured line of sight to this source, which should therefore be switched off, leaving only reflection from distant material as an echo of its past activity.

Alternatively, assuming that the normalization of the intrinsic continuum is the same as the scattered one, the column density along the line of sight must be $>2\times10^{24}$ cm$^{-2}$, which is significantly larger than the column density along the line of sight derived from the default \textsc{MyTORUS} geometry (roughly between 0.25 and 0.65 times the equatorial one, given the constraints we get on the inclination angle), and than the mean column density, integrated over all lines-of-sight through the torus, which is $\pi/4$ the equatorial one. Therefore, in this scenario, the absorbing/reflecting medium would not be homogeneous, but the column density along the line of sight, responsible for the absorption of the nuclear continuum, would be larger than the average column density of the reflector, responsible instead of the overall properties of the reflected spectrum. The 14-195 keV {\it Swift}-BAT spectrum of ESO 138-G1 retrieved from the online 70-month catalog\footnote{\tt http://swift.gsfc.nasa.gov/results/bs70mon/} does not allow us to discriminate between the two scenarios due to its low statistics. We also note that the galaxy NGC 6221 (only $\sim 11$ arcmin distant from ESO 138-G1) might possibly contaminate the {\it Swift}-BAT FOV.
A possible breakthrough can be represented by observations preformed with high energy ($>$10 keV) focusing instruments, such {\it NuSTAR} and, in the near future, {\it Astro-H}, which could allow us to disentangle the hard X-ray spectral properties of these different scenarios and discard the contamination due to the presence of NGC 6221 in the FOV.

\section{Conclusions}
We presented in this work a broadband (0.3-10 keV) analysis of the XMM-{\it Newton} and {\it Chandra} observations of the Seyfert 2 galaxy, ESO 138-G001. Our main results can be summarized as follows.

\begin{itemize}
\item The source appears to be absorbed by Compton-thick gas along the line of sight, confirming and refining past results. However, if a self-consistent model of reprocessing from cold toroidal material is used (\textsc{MyTORUS}), a possible scenario requires the absorber to be inhomogenous, its column density along the line of sight being larger than the average column density integrated over all lines-of-sight through the torus;\\

\item the iron emission line around 6.4 keV is likely produced by moderately ionised iron (FeXII-FeXIII), as suggested by the shifted centroid energy and the low K$\beta$/K$\alpha$ flux ratio;\\

\item the soft X-ray emission is dominated by emission features, well identified in the high-spectral resolution RGS spectra with transitions from highly ionised metals, as commonly found in Seyfert 2 galaxies. This emission is likely associated with the optical NLR, which is quite compact and, therefore, unresolved even in the high-spatial-resolution \textit{Chandra} image. Photoionisation appears to be the main excitation mechanism for this gas, as confirmed by line diagnostics and the use of self-consistent models (\textsc{Cloudy}) for the low-resolution pn spectrum.

\end{itemize}

\section*{Acknowledgments}
We thank M. Guainazzi for useful discussions.
\bibliographystyle{mn2e}
\bibliography{sbs} 
\end{document}